\documentclass[
      ,draft            
  ]
  {aipproc}

\layoutstyle{6x9}

\begin{document}

\title{Constraints on the Diverse Progenitors of GRBs from the Large-Scale Environments}

\classification{98.70.Rz}

\author{J. S. Bloom}{
  address={Astronomy Department, University of California, Berkeley, 601 Campbell Hall, Berkeley, CA 94720}
}

\author{J. X. Prochaska}{
  address={Department of Astronomy and Astrophysics, UCO/Lick Observatory; University of California, 1156 High Street, Santa Cruz, CA 95064}
}

\begin{abstract}
 
The pursuit of the progenitors of short duration-hard spectrum gamma-ray bursts (SHBs) draws strongly upon similar quests for the origin of supernovae (SNe) and long duration-soft spectrum GRBs (LSBs). Indeed the notion that, in the absence of smoking guns, the progenitors of cosmic explosions betray their identities both on the global and local scale, motivates the study of SHB redshifts, host galaxies, and locations with respect to hosts. To this end, we suggest both a historical and emergent physical analogy of GRBs with SNe: long-soft GRBs are to core-collapsed supernovae as short-hard GRBs are to Type Ia supernovae (``LSB:CC::SHB:Ia''). 
Still, the SHB progenitor pursuit is just beginning and we caution that while there are some substantive differences between observations of LSBs and SHBs on large-scales, particularly in host demographics, neither the offset nor the redshift distributions of SHBs are statistically inconsistent with those of LSBs.
\end{abstract}

\maketitle


\section{Introduction}

The subclassification of high-energy bursting phenomena relies, for good reason, on the properties of the prompt emission. Yet whether the diversity of burst duration and spectral hardness 
reflects a true diversity in the physics of the explosions and the nature of the progenitors, was largely a leap of faith for the vast majority of GRBs. Soft-gamma ray repeaters aside, there was until 2005 little evidence to suggest a real difference between long-soft (LSB) and short-hard (SHB) GRBs \cite{Kou93}. Following the first rapid locations of SHBs by 
Swift \cite{Geh05} and {\it HETE-2} \cite{Vil05}, the first X-ray afterglows were found \cite{Geh05}, allowing for an unprecedented localization of five SHBs. K.\ Hurley gave an excellent and thorough overview of high-energy properties of short-hard bursts in this conference.

What has emerged with the limited set of SHBs are both commonalities and striking differences between the two populations. Broadband modeling \cite{bpp+06,ffp+05,pana05,sbk+06} has shown a broad consistency of afterglows with the external shock model.  As predicted \cite{pkn01}, because the luminosity of the afterglow scales with the input energy of the explosion, the afterglows of SHBs appear to be fainter analogues of LSBs.  Moreover, there is now evidence that short bursts are weakly collimated with half angle jetting of more than 10\% (see \cite{sbk+06}). Yet unlike with the relatively co\"operative LSBs, it may be some time before a definitive observation reveals the origin of short bursts: the best smoking gun for the merger hypthothesis is a concident detection of a burst of gravitational waves.

Thankfully, the progenitors of cosmic explosions are observationally manifested on length scales ranging from circumburst to galaxy cluster. Indeed, a census of the {\em types of host galaxies} of SNe beginning in the 1950s yielded insights into the various progenitors: Reaves \cite{rea53} noticed that Type I (now subclassed as Ia) SNe occurred in galaxies of all sorts whereas Type II arise from essentially only in late-types. This observation is fully borne out in modern studies \cite{ctt+97}. Likewise, systematic surveys of the {\it locations of SNe around galaxies} further informed the progenitor question. Van Dyk \cite{vd92} concluded  that the location of SNs ``suggest that type Ia supernovae do not arise from massive short-lived stellar populations. Type Ib/Ic and type II supernovae, however, are very likely to be associated with H {\sc II} regions and therefore with massive stellar progenitors.'' Recently, measurements of the {\it redshift distribution} of Ia SNe have shown evidence for delay since starburst \cite{srd+04}. All such observations of SNe on large scales support the established belief from detailed modeling of the transient emission that massive young stars are responsible for Type II SNe and degenerate old stars are responsible for Ia SNe.

As reviewed below, the quest for an understanding of the progenitors of LSBs benefited from a similar analysis of observations on large scales. Now, thanks to the breakthrough localization of afterglows and motivated by past work with SNe and LSBs, the search for the origin of SHBs is following parallel (productive) tracks.

\section{Progenitors Revealed on Large Scales}

\begin{table}
\begin{tabular}{lrrrrrr}
\hline
  & \tablehead{1}{r}{b}{redshift}
  & \tablehead{1}{r}{b}{host\\classification}
  & \tablehead{1}{r}{b}{in\\cluster?}
  & \tablehead{1}{r}{b}{SFR\\ ($M_\odot$ yr$^{-1}$)}  
  & \tablehead{1}{r}{b}{Min Age\\ Since Starburst\\ (Gyr)}
  & \tablehead{1}{r}{b}{Metallicity\\ ($Z/Z_\odot$)}   \\
\hline
050509b & 0.225 & E & Yes & $<$0.1 & $\sim$1 & $\sim$1 \\
050709  & 0.160 & Irr/late-type dwarf & No  & $>0.3$ & $\sim$ongoing & 0.25 \\
050724  & 0.258 & early (E+S0) & No & $ < 0.05$ & 8 & 0.2 \\
050813  & 0.722 & E?    & Likely & $<$0.2  & ~ & $\sim$1\\
        & (1.7?)\tablenote{Berger, this conference} \\
051221  & 0.5459& late-type dwarf? & ? & $\sim$0.2  & ~ & $\sim$1\\
\hline
\end{tabular}
\caption{
Basic Properties of Short-Hard GRB Hosts
  }
\source{\cite{bpp+06,Geh05,bpc+05,pbc+05,ffp+05,bcb+05,sbk+06,gcg+06} and Covino (this conference)}

\label{tab:hosts}
\end{table}

\subsection{Host Galaxies}

The population of LSB hosts appear to be best classified as ``faint blue galaxies." (see reviews by N.\ Tanvir and E. Le Floc'h, this conference). To summarize, they are underluminous (median $L \sim 0.1 L_*$) with disturbed/irregular morphologies. Indeed, of the more than 50 LSBs that have been well-localized to date, no LSB has been definitively associated with an early-type host. The hosts do appear to be especially ripe environments for making collapsars \cite{sfc+01,dkb+01,Fyn03,ldm+03,chg04,cvf+05,wbp05,fru+05}. In particular, the star formation per unit mass appears high --- with half of well-studied hosts showing more than 10 $M_\odot$ yr$^{-1}$ $L_*^{-1}$ --- indicating recent star burst activity (\cite{vfk+01,Fyn03,chg04,jbf+05} and Prochaska, this meeting). There is also anecdotal evidence that the ratio of [Ne III]/[O II] is high, indicative of hot H II regions, and suggesting a propensity for making massive stars. While the hosts do appear to be low in metallicity, it is not clear whether the hosts are, in general, lower in metallicity than galaxies at similar redshifts.


In contrast to the LSB hosts,  the galaxies putatively associated
with the SHB population exhibit a wide range of 
morphology, star-formation rate, and metallicity.  The
SHB host population currently includes both solar metallicity
elliptical galaxies and metal-poor late-type dwarfs \cite{pbc+05,sbk+06}.  
The discovery of the early-type galaxies hosts argue for a greater than 1~Gyr
delay between the formation of the progenitor star(s) and
the burst.  It may also be notable that even the putative hosts with
current star-formation exhibit evidence for old stars
(Covino, this conference; \cite{sbk+06}), although this
is a common characteristic of dwarf galaxies in the Local Group (e.g., \cite{scgs02}).

If SHBs share a common progenitor, it may be possible to constrain
the lifetime of SHB progenitors based on the morphological mix
of host galaxies (\cite{gno+06}; Ramirez-Ruiz, this conference).
For example, the morphological types for the first four bursts in
Table~2 reflect a higher incidence of early-type galaxies than
Type~Ia SN  and suggest a progenitor lifetime significantly exceeding
1~Gyr.  These conclusions, however, are not  robust and are subject to the small
size of the current sample (for example, the inclusion of GRB~051221 found after the conference increased
the late-type fraction by a factor of two!).  On the other hand, 
a large progenitor lifetime would help explain the apparent
high incidence of cluster membership.  This could be naturally
explained by the fact that galaxies in overdense regions form
earlier in hierarchical cosmologies \cite{gno+06}.

\subsection{Redshift Distributions}


If LSBs arise from the death of massive stars, then a natural ramification is that the long-soft bursting rate should trace the universal star formation rate (SFR). The brightness of the gamma-ray data alone has shown broad consistency (e.g., \cite{lw98}) but it was not until relatively recently, with the advent of spectroscopic redshifts of LSBs, that details of the bursting rate could be inferred. The difficultly in making the comparison is two-fold. First, the ``true" universal star formation rate is not known a priori and is subject to observational selections such as the suppression of UV starlight due to dust. The canonical model rate used for GRB litmus tests, parameterized in Porciani and Madau \cite{pm01}, has the SFR rising rapidly from $z=0$ to $z=1$, then flat out to large redshifts. Second, there are inherent biases in redshift discovery of GRBs which are often difficult to quantify, such as translating non-detections of spectroscopic redshifts into limits and using complex trigger efficiencies near the detection level to infer rates at the faint end of the flux distribution. Redshifts found mostly from emission lines  in the host -- the majority of pre-Swift redshifts -- are especially subject to complex biases due to the finite bandpass of optical spectrographs, emission from night sky lines, and limited number of SF emission lines; nevertheless, pre-Swift LSBs were shown consistent with the SFR (e.g.\ \cite{blo03a}). Redshifts discovered by means of absorption-line spectroscopy, how most Swift burst redshifts are found thanks to rapid localizations, are significantly less subject to observational biases. P.\ Jakobsson and B.\ Gendre presented results on the true redshift rate in the Swift era, finding a reasonable consistency with the universal SFR (see also \cite{jlf+05}).

\begin{figure}
  \includegraphics[width=.95\textwidth,angle=0]{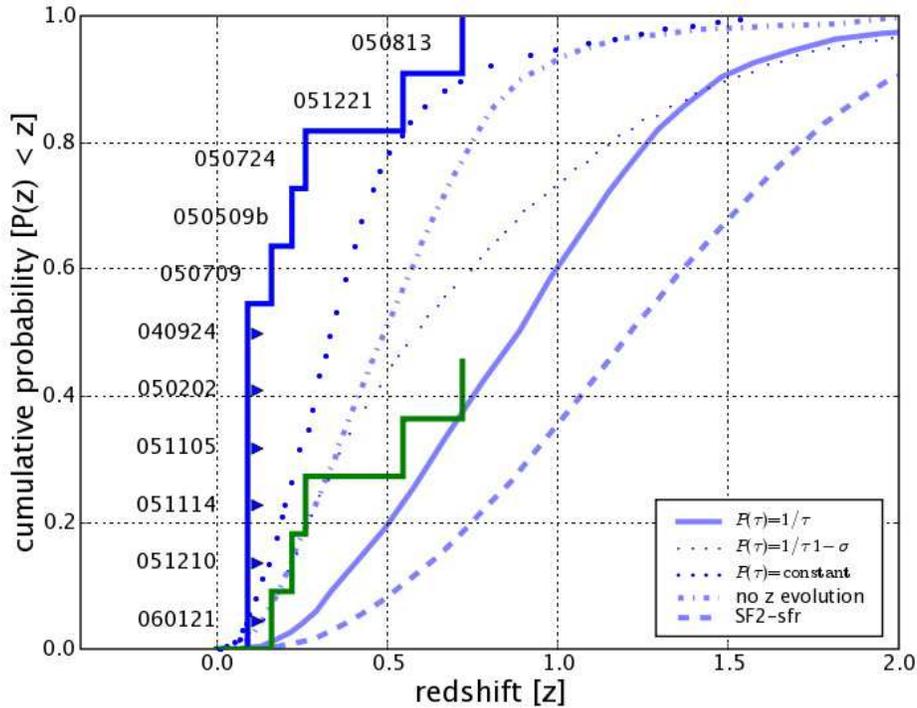}
\caption{A ``complete'' redshift distribution of those SHBs localized rapidly by {\it HETE-2} and Swift, not associated with SGRs, compared with progenitor models (from \cite{gp05}) with different temporal delays from starburst. As of 21 Jan 2006, there are 6 SHBs without redshift (shown in the leftmost cumulative histogram with right-pointing arrows) and 5 with redshift in this sample. The truncated histogram at right, shows how the first 5 SHBs with redshift would lie if all those SHBs without redshift arose from $z$ greater than 0.7. Aside from tracing the universal SFR (SF2, dashed line), the current sample does not exclude any of the Guetta \& Piran models. If a redshift for SHB 050813 of $z=1.7$ is adopted (Berger, this conference), then the SF2 model is statistically allowed by the data.}
  \label{fig:z}
\end{figure}


As the individual {\em a posteriori} identification of the 5 SHBs with afterglows with specific redshifts range from $P \approx 10^{-4}$ (e.g., 050904) to $P \approx 10^{-3}$ (e.g., 050509b), there is now little doubt that the population short-hard GRBs are of extragalactic origin.
Still, as of writing, there has been no absorption line redshift measured for a {\em bone fide} short-hard GRB: the inference of SHB redshifts relies at present on the statistical connection to a putative host galaxy and spectroscopy of that host. 

A cursory comparison of the redshift distribution of SHBs with LHBs reveals what appears to be a significantly different population. Augmenting those  5 SHBs (with known afterglows) in Table \ref{tab:hosts} with 790613 ($z = 0.09$, inferred by statistical arguments of burst-only position \cite{gno+06}), the population indeed appears to be heavily skewed to lower redshifts. Whereas the median redshift of LSBs is $z = 2.8$ \cite{jlf+05}, of these 6 reasonably secure redshifts of SHBs, the median is $z = 0.24$. Guetta \& Piran \cite{gp05} and Nakar et al.~\cite{ngf05} have suggested that the difference could be only partly due to the relative faintness of the prompt burst emission brightness resulting in only the relatively nearby SHBs causing a trigger.   Despite the brightness bias, the Guetta \& Piran claim is that the intrinsic rate of SHBs are more skewed to lower $z$ than the universal SFR. Nakar et al.\ further claimed (before the $z=1.7$ hypothesis by E.~Berger was advanced at the conference and 051221 was discovered) that the true SHB redshift distribution requires a progenitor population which goes boom more than a few Gyr after starburst. If true, this stretches even the longest-timescale merger scenarios.

Given the instrumental and environmental biases against detection of higher redshift SHBs, the non-detections of SHB redshifts in a complete sample must be properly accounted for statistically when drawing conclusions about the true rate. As an illustration, if we define the sample as ``all {\it HETE-2} and Swift short-hard bursts which were promptly localized but not associated with an SGR'' the list includes those in Table \ref{tab:hosts} plus GRBs 060121, 051210, 051114, 051105, 050202, and 040924; that is, 5 with redshifts and 6 without\footnote{At the conference, we showed similar results using small IPN error boxes as the defining sample of SHBs}. Applying survival analysis on the sample\footnote{Since the error circles of those without redshift were searched for nearby galaxies to depths exceeding the DSS limit, I have assumed a lower limit of $z=0.09$ (following Gal-Yam et al.\ on IPN bursts).}, we find that the data are consistent with having been drawn (and censored) at random by all of the temporal delay models proposed by Guetta \& Piran. Even the universal star-formation rate (SF2) is allowed (e.g., under the Peto \& Peto Generalized Wilcoxon Test) with a few percent chance. The Kaplan Meier mean is $z = 0.38 \pm 0.10$. If SHB 050813 arose from $z=1.7$ (as suggested by Berger at this conference) the Kaplan Meier mean redshift is $z$ = 0.58 $\pm$ 0.26 and the data are consistent statistically with the SF2 model. The conclusions, which implicitly ignore all the short bursts too faint and distant to trigger on prompt emission, are clearly still sensitive to individual bursts; this reflects the danger of drawing strong conclusions from the current small sample.

\subsection{Locations In and Around Galaxies}
\begin{figure}
  \includegraphics[width=.95\textwidth,angle=0]{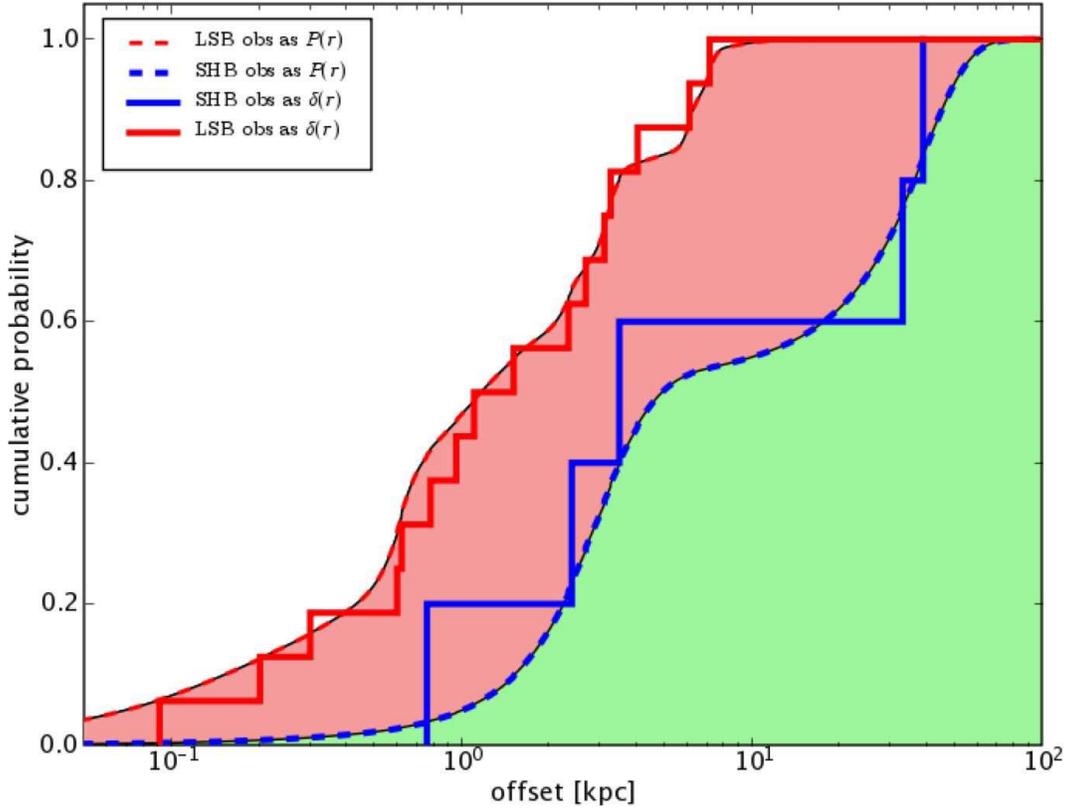}
\caption{Comparison of the cumulative offset distribution of long-soft (left) and short-hard GRBs (right). The SHB sample is taken from Table \ref{tab:off} and the LSB sample comprises the first 16 bursts with known redshifts and offsets. The histograms are made assuming that the offsets are known precisely (ie., a $\delta$-function at the measured offset), whereas the smooth curves account for the uncertainty in the offset measurements following the formalism of \cite{bkd02}. While SHBs {\em appear} qualitatively to be more diffusely located with respect to their putative hosts, the Kolomogorov-Smirnoff probability that the observed  SHB population was drawn at random from the observed LSB population is 38\%. That is, the data do not yet support the qualitatively assessment that the locations of SHBs are substaintially different than LSBs. }
\label{fig:off}
\end{figure}

From the first sub-arcsec localization of LSBs, showing an offset of GRB\, 970228 from the centroid of the optical light (e.g. \cite{slp+97}), it was clear that the phenomena was unlikely related to activity of the central black hole \cite{car92}. Unfortunately, due to the relatively large redshifts of LSBs, resolved imaging of the immediate environment of bursts (say at the 100 pc level) is simply not possible even with diffraction limited imaging of HST and 10-m groundbased telescopes. Inferences about what the locations of GRBs imply for the progenitors must be made in a statistical manner. A comparison of the first 20 GRB localizations with apparent centers of their hosts revealed two surprising results in the context of the then paradigm of LSBs as merger products. First, the locations are relatively concentrated towards the centers of galaxies, all less than $10$ kpc from the apparent centers.  This tight concentration is statistically inconsistent with nominal merger locations. [The standard NS-NS formation scenarios usually lead to median merger times between 0.1 -- 1 Gyr from starburst with systemic kick velocities of order 100 km s$^{-1}$. The expected offsets from mergers is of course dependent on the host properties\footnote{Since the hosts are relatively low mass, suggesting that double NSs could escape the host potenitial before coalescence, a tight concentration is especially disfavored. This inconsistency is entirely dependent upon the dominant channel for double NS production. Other production channels do posit small offsets of GRBs from star formation locations (e.g., \cite{pb02}).}.] Second, the locations of LSBs appeared to trace the location of the UV light of hosts, suggesting an intimate connection of LSBs with star formation. A.\ Fruchter presented results at the conference, making use of HST imaging of new GRBs, and showing a continued connection of LSBs with the light of their hosts. In particular, the locations of LSBs appear to prefer some of the highest surface brightness components of their hosts. 

\begin{table}
\begin{tabular}{lrr}
\hline
  & \tablehead{1}{r}{b}{Angular Offset\\ (arcsec)}
  & \tablehead{1}{r}{b}{Projected\\ (kpc)}\\
  \hline
050509B & 10.9 $\pm$ 3.6  & 39 $\pm$ 13 \\
050709  & 1.29 $\pm$ 0.48 &  3.5 $\pm$ 1.3 \\
050724  & 0.61 $\pm$ 0.23 &  2.4 $\pm$ 0.9 \\
050813\tablenote{Here, we have assumed that source "B" in Prochaska et al.\ is the host. This is by no means the only possible offset for this burst. The deprojection is made assuming $H_0$ = 71 km s$^{-1}$ Mpc$^{-1}$, $\Omega_m = 0.3$, $\Omega_\Lambda = 0.7$} & $4.66 \pm 2.5$ & 33.2 $\pm$ 17.6 \\
051221  & 0.12 $\pm$ 0.04 & 0.760 $\pm$ 0.030 \\
\hline
\end{tabular}
\caption{Offsets of Short Gamma-Ray Bursts from Their Putative Host Galaxies
  }
  \source{\cite{pbc+05,sbk+06}}
\label{tab:off}
\end{table}

In light of the LSB location results, one of the startling observations of SHB 050509b was not only the statistical connection to a low-redshift giant elliptical galaxy, but, if the connection were true, the relatively large offset from that galaxy. Whereas no LSB had been found more than 10 kpc in projection from a putative host, the first SHB was well-localized at 39 $\pm$ 13 kpc in projection. This, in and of itself, is suggestive of a different population of progenitors for SHBs. In contrast, the location of SHB 050709, on the outskirts of a faint star-burst galaxy bears a striking resemblance to the burst-host configuration of LSB 970228.  As with LSBs, analysis of {\em individual} locations would clearly lead to diverse conclusions about the nature of the progenitors. 

Thankfully, while there are only 5 SHBs with $\sim$arcsec localizations (see Table \ref{tab:off}), we are now in the position to  begin to test various progenitor scenarios\footnote{It is a popular misconception that NS--NS merger locations are unlikely to be close (less than $\sim$5 kpc) to their hosts based on the cumulative offset distribution presented in Figure 22 of Fryer, Woosley, Hartmann \cite{fwh99}. It appears that the scale of the x (distance) axis in that figure is incorrect by an order of magnitude with respect the differential distributions shown in Figure 21. To be sure, the median offset for an $M_*$ galaxy is $\sim$10 kpc, which is consistent with other simulations from other groups.}. What can be stated with certainty is SHBs do not all arise from the central activity of galaxies. Following Figure \ref{fig:off} SHBs {\it appear} to be more diffusely positioned around galaxies than LSBs. But, owing to the small numbers and sometimes large uncertainties in the offset measurements, we find that the locations of SHBs are still statistically significant ($P_{\rm K-S} = 0.38$) with having been drawn from the same population of as LSBs. 

\section{Discussions and Conclusions}

The search for the origin(s) of SHBs presents a substantial challenge relative to discovery of long-burst progenitors: there is no obvious smoking gun for short burst progenitors that have not already been looked for. Whereas an obvious manifestation of the collapsar model was the connection to star forming hosts and star forming regions within hosts, viable progenitor models predict offsets and locations within hosts that are far from robust. So what may we conclude about how the current host observations of SHBs relative to LSBs reflect the progenitors? A secure statement is that "SHB hosts contain a generally older population of stars than LSB" and a reasonable, but somewhat more directed statement is that "SHBs come from an old stellar population whereas LSBs do not." What we think is still a matter of debate is whether ``the frequency of early-type to late-type hosts for SHB is consistent with NS--NS" (E.\ Ramirez-Ruiz discussed his views on this in the conference). Similar statements can be made of the offset distribution of SHBs: yes, there appears to be some differences between LSBs and SHBs but nothing stastistically significant at this time. What appears relatively secure, based on the small offsets of three of 5 SHBs from low-mass galaxies is that:
\begin{quote}
 {\bf The progenitors of SHBs cannot have both large systematic kicks ($> 100$ km s$^{-1}$) at formation AND inhere large delay times from starburst ($> 1$ Gyr)}.
\end{quote}

\noindent Making even stronger statements about the nature of the progenitors is not only hampered by small number statistics but in the lack of robust predictions from the models. But the above statement can be reworked in the form of a generic set of predictions:

  \begin{itemize}
    \item In Long delay ($>$1 Gyr) progenitors scenarios with kicks the offsets should anti-correlate with host mass and correlate with average stellar age 
    \item In Short delay  ($<$ 1 Gyr) progenitors the offsets should correlate with host mass and anti-correlate with average stellar age.
   \end{itemize}
   
\noindent If the progenitor lifetime of the SHBs is long and kicks
are small, then the bursts should correspond spatially to
the oldest stellar populations in a given galaxies.  For
early-type galaxies, the distribution would presumably follow
the light of the galaxy.  In contrast, the distribution
in star-forming galaxies might be more concentrated in the
spheroid (e.g., bulge of the Milky Way).  

Those of us that have worked on creating {\it ab initio} predictions for NS--NS progenitors all basically agree on the offsets and delays provided we use a production channel dominated by binaries where the helium stars do not fill their respective Roche lobe before exploding as a SN. But other productions might dominate, leading to predictions of relatively short merger times and small offsets \cite{pb02,bpb+06}. Both T.\ Piran and J.\ Grindlay discussed other production channels in the conference. For example, if SHBs arise from NS binaries made in the centers of globular clusters \cite{gpm06}, then the offsets should scale with the halo size of the host (rather than exponential disk size) and the systemic kicks could be small. This results in diverse predictions of offsets and host demography for the same progenitors.

Fostered by the presence of a curious 100 sec timescale for prompt emission and lacking the detection of a Li-Paczy\'nski minisupernova \cite{lp98b}, the theory community is rapidly developing other progenitor models for SHBs such as WD--WD mergers \cite{lwc+06} and accretion induced collapse of a NS \cite{da06,mrz06}. In this respect the connection of GRBs with SNe may not only be a historical one, there may be a deeper analogy: long-soft GRBs are to core-collapased supernovae as short-hard GRBs are to Type Ia supernovae (``LSB:CC::SHB:Ia''). Not only does this ring true for the hosts and locations (like Type Ia SNe, short bursts can occur outside star forming regions in galaxies of any type, whereas, like core-collapsed SNe, long bursts only occur near star forming regions of galaxies actively forming stars) but could also reflect a symmetry of the progenitors: LSBs result from a special (extreme?) version of core-collapased massive young stars whereas SHBs might arise in the NS to BH transition (and Ia SNs in WD to NS transition).

Whether the LSB:CC::SHB:Ia analogy will hold at the physical level remains to be seen. But as progenitor predictions for SHBs are honed, new SHB offsets plus inferences of the host mass will continue to constrain.  Still, a direct comparison with model predictions will also be hampered by the possibly of an ``ambient density bias," where afterglows, the brightnesses of which scale as the circumburst density to the half power, are more likely to be discovered if the burst occurs in regions of larger density.  Are we missing a population of bursts at large offsets that occur in the intergalactic medium? If the internal shock paradigm for short burst holds then $\sim$arcsecond location of the prompt emission in future missions should be immune to this bias.
Last, we note that while SHB afterglow follow-up tends to be more difficult than LHBs, one positive result of the low $z$ nature of detected SHBs is that we may have the possibility of resolving the sub-structure of the birthsites within galaxies (e.g., spiral arms, H {\sc II} regions) using HST and groundbased adaptive optics.




\begin{theacknowledgments}

We are thankful to Daniel Perley, Enrico Ramirez-Ruiz, Jonathan Granot, and Hsiao-Wen Chen for their respective insights and work on the nature of short bursts. JSB and JXP acknowledge support from NASA/Swift grant NNG05GF55G. JSB thanks the organizing committee for partial support of students Perley and Katey Alatalo.

\end{theacknowledgments}





\IfFileExists{\jobname.bbl}{}
 {\typeout{}
  \typeout{******************************************}
  \typeout{** Please run "bibtex \jobname" to optain}
  \typeout{** the bibliography and then re-run LaTeX}
  \typeout{** twice to fix the references!}
  \typeout{******************************************}
  \typeout{}
 }

\end{document}